\documentclass[12pt]{article}
\def\beq{\begin{eqnarray}}
\def\eeq{\end{eqnarray}}
\def\beqs{\begin{eqnarray*}}
\def\eeqs{\end{eqnarray*}}

\newcommand{\be}{\begin{equation}}
\newcommand{\ee}{\end{equation}}
\newcommand{\lll}{\langle}
\newcommand{\rrr}{\rangle}

\newcommand{\T}{\mbox{Tr}\> }
\newcommand{\rir}{\right\rangle}
\newcommand{\lel}{\left\langle}
\def\centeron#1#2{{\setbox0=\hbox{#1}\setbox1=\hbox{#2}\ifdim
\wd1>\wd0\kern.5\wd1\kern-.5\wd0\fi
\copy0\kern-.5\wd0\kern-.5\wd1\copy1\ifdim\wd0>\wd1
\kern.5\wd0\kern-.5\wd1\fi}}
\def\ltap{\;\centeron{\raise.35ex\hbox{$<$}}{\lower.65ex\hbox{$\sim$}}\;}
\def\gtap{\;\centeron{\raise.35ex\hbox{$>$}}{\lower.65ex\hbox{$\sim$}}\;}

\begin{document}
\begin{titlepage}
\begin{flushright}
{ITP-UU-02/48}
\end{flushright}

\vskip 1.2cm

\begin{center}

{\LARGE\bf On the pattern of Casimir scaling violation in
gluodynamics }

\vskip 1.4cm

{\large  V.I. Shevchenko}
\\
\vskip 0.3cm
{\it Institute for Theoretical Physics, Utrecht University,
Leuvenlaan 4 \\ 3584 CE Utrecht, the Netherlands}
\\
\vskip 0.1cm
and \\
\vskip 0.1cm
{\it Institute of Theoretical and Experimental Physics,
B.Cheremushkinskaya 25 \\ 117218 Moscow, Russia } \\
\vskip 0.25cm
e-mail: V.Shevchenko@phys.uu.nl

\vskip 2cm

\begin{abstract}
Results of lattice analysis indicate that the static potential in
$SU(3)$ gauge theory is proportional to eigenvalue of quadratic
Casimir operator for the corresponding representation with a good
accuracy. We discuss the pattern of deviation from this Casimir
scaling in gluodynamics in terms of correlators of path-ordered
gauge-invariant operators defined on the worldsheet of the
confining string.
\end{abstract}
\end{center}

\vskip 1.0 cm

\end{titlepage}

%%%%%%%%%%%%%%%%%%%%%%%%%%%%%%%%%%%%%
% COUNTERS
%%%%%%%%%%%%%%%%%%%%%%%%%%%%%%%%%%%%%
\setcounter{footnote}{0} \setcounter{page}{2}
\setcounter{section}{0} \setcounter{subsection}{0}
\setcounter{subsubsection}{0}

%%%%%%%%%%%%%%%%%%%%%%%%%%%%%%%%%%%%%%%%%
% THE MAIN TEXT
%%%%%%%%%%%%%%%%%%%%%%%%%%%%%%%%%%%%%%%%%

\section{Introduction}

Physics of confinement in QCD has been attracting considerable
attention since the Yang-Mills theory with dynamical quarks was
established as a theory of strong interactions. Despite
significant theoretical progress the operational framework for the
exact calculations beyond perturbation theory has not been found,
leaving aside numerical simulations of the theory on the lattice.
In particular, one has no clear understanding of the relation
between confinement and mass gap property. It is commonly believed
that confining theory such as gluodynamics (i.e. QCD with no
quarks) exhibits the mass gap, e.g. the lightest excitation over
the physical vacuum is a massive particle. The latter one is
indicated by the lattice calculations to be scalar $0^{++}$
glueball with the mass $m\approx 1.5 $ Gev in the case of $N=3$.
On the other hand, if one studies confining potential between
static sources in pure Yang-Mills theory, the physics of area law
is governed by the formation of the confining string with nonzero
tension $\sqrt{\sigma} \approx 0.4 $ Gev and this process is not
directly related to the details of glueball physics. The dynamical
reasons for that are well known -- corresponding effective
interaction is suppressed by relatively large mass of the
glueballs in the units of $\sqrt{\sigma}$ and also by
$1/N$--factor. Put in simple terms, even if one acquired full
knowledge of the glueball spectrum in gluodynamics, it would not
help to compute generic Wilson loop. It is worth mentioning that
in the abelian Higgs model with electrically charged condensate
the mass of the photon (which plays the role of the lightest
glueball in this theory) enters monopole-antimonopole string
tension in rather nontrivial way and in combination with other
quantities, giving the simplest example of the same nature.

The standard approach of quantum field theory describes
interactions in terms of particle exchanges. In confining theories
the applicability of this program is limited, in some sense:
string creation, which is one of the most important manifestations
of confinement can be understood neither in terms of gluon
exchanges nor as a result of glueball interactions.\footnote{We do
not discuss the models with the so called "confining gluon
propagator" since they typically miss the phenomenon of
gauge-invariant confining string formation as well.} Nevertheless
it is desirable to establish some connection between different
model approaches or at least to understand, where such relations
are to be expected. The aim of the present paper is to make a step
in this direction. It will be shown that the certain linear
combination of the static potentials arising from the Wilson loops
in different representations gets contributions from the exchange
by colorless states. Our actual analysis will be presented for
adjoint and fundamental loops which are of the main
phenomenological interest but we believe that the mechanism is
quite general. The paper is organized as follows: theoretical and
lattice status of static potentials in different representations
is briefly discussed in the Section 2, the Section 3 presents the
details and results of our analysis on more technical level, and
short conclusion can be found in the Section 4.

\section{Static potentials }

Recent lattice results on static potentials in different
representations for $SU(N)$ Yang-Mills theory
\cite{bali1,bali2,deldar1,deldar2} have attracted considerable
interest (see \cite{bali2} for references to early studies). It
was argued in \cite{sim1,ss1,ss2} that the data can serve as a
good test for different phenomenological descriptions of
confinement in gluodynamics and QCD. The reason for that lies in
the impressive agreement between the lattice data and the so
called {\it Casimir scaling} (CS) hypothesis \cite{amb} which
states\footnote{The term {\it Casimir scaling} was proposed in
\cite{ddil}} proportionality of representation--dependent
confining potential (and, in more restricted form, of the string
tension) to an eigenvalue $C_D$ of quadratic Casimir operator
${\cal C}^{(2)}_D$ for the given representation $D$. It is
interesting that many microscopic models of confinement meet
serious difficulties in reproducing CS (see discussion in
\cite{ss2}).

However, one is not to expect CS to be exact law in the case under
consideration. As a matter of principle, CS should be violated at
large distances because of screening. Qualitative picture for
asymptotically large Wilson loops corresponds to the string
tension $\sigma_D$ equal to fundamental string tension $\sigma =
\sigma_{fund}$, if the Wilson loop in the representation $D$
nontrivially transforms under the center of the gauge group, or
zero otherwise, in obvious contradiction with CS. On the other
hand, the actual value of critical distance where effects of
screening become important may be as large as 1.4 Fm \cite{ss2}
and one still has large enough "CS window", where the corrections
to CS due to screening are exponentially small. As to perturbation
theory, CS law for static potential is known to hold up to two
loops in continuum theory \cite{psr,psr1}, however there is no any
general theorem that CS should be exact property of perturbative
potential and analysis of diagrams provides qualitative reasons in
favor of CS violation at the three-loop level.
 It is also worth mentioning that CS is violated in the strong
coupling expansion at the order $1/g^8$ \cite{deldeb2} while it
holds at lower orders.

Another line of research confronts CS with alternative theoretical
predictions for $D$--de\-pen\-dence of the static potential. The
only law different from CS which has not yet been definitely ruled
out in lattice simulations is the so called $sin$-formula,
motivated by supersymmetric Yang-Mills theory analysis
\cite{douglas,strassler}. Moreover, the authors of \cite{deldeb1}
claim that their numerical data for asymptotic string tension in
$d=3+1$ nonsupersymmetric $SU(N)$ theory with $N=4,6$ favor
$sin$-formula and disagree with CS (see also \cite{gow}). On the
contrary, authors of \cite{lt} claim their results to be in
agreement with both CS and $sin$-formula predictions within two
standard deviations in $d=3+1$ Yang-Mills theory and definitely
deviate from both in $d=2+1$ case. It should be noted that in all
studied cases the numerical difference between CS law and
$sin$-formula is actually quite small that plagues the
discrimination between the two predictions.

Unfortunately, as such, the $sin$-formula gives no nontrivial
testable predictions for physically interesting case of the gauge
group $SU(3)$. On the other hand, there are theoretical reasons to
believe \cite{konishi} that it receives non-universal corrections.
This seems to be a good motivation to take the predictions like CS
or $sin$-formula as the first approximations and study the pattern
of possible deviations from these regimes. Needless to say that
despite numerically the two predictions are surprisingly close,
the qualitative physical picture behind them is drastically
different. In any case, data from
\cite{bali1,bali2,deldar1,deldar2} as well as from
\cite{deldeb1,gow,lt} undoubtedly indicate that CS as the first
approximation for static potential works with percent-level
accuracy at not too large distances.

The starting point is the general expression for the Wilson loop
\be \lll W_D(C)\rrr = \left\lll {\T}_D \>{\mbox
P}\exp\left(i\int\limits_C A_{\mu}^a T_D^a
dz_{\mu}\right)\;\right\rrr \label{wloopap} \ee where the
generators $T_D^a$ correspond to irreducible representation of
dimension $D$. The normalized trace ${\T}_D$ is defined as ${\T}_D
{\hat1}_D = \frac{1}{D}  \T {\hat 1}_D = 1$, fundamental and
adjoint generators are normalized according to \be\T T^a_{fund}
T^b_{fund} = {\delta}^{ab} / 2 \;\;\; ; \;\;\; \T T^a_{adj}
T^b_{adj} = N {\delta}^{ab} \ee The eigenvalue $C_D$ of quadratic
Casimir operator ${\cal C}^{(2)}_D $ is defined as follows \be
{\cal C}^{(2)}_D = \delta_{ab}\> T_D^a T_D^b = T_D^a T_D^a =
C_D\cdot {\hat1}_D \label{casimir} \ee We will confine our
attention to adjoint representation in the present paper. The
corresponding Casimir ratio $d_{D_2 / D_1} = C_{D_2} / C_{D_1}$ is
given by $d_{a/f} = 2N^2 / (N^2 -1)$, where the fundamental
Casimir $C_{fund} = (N^2-1)/2N$ and adjoint one $C_{adj} = N$
equal to $4/3$ and $3$, respectively, for the case of $SU(3)$. The
static potential is formally defined as \be V_D(\mu, R) = -
\lim_{T\to\infty} \> \frac{1}{T} \> \log \lll W_D(C)\rrr
\label{potent1} \ee where $\mu $ stays for renormalization
scale.\footnote{On the lattice the role of $\mu$ is played by
inverse lattice spacing $a^{-1}$} The physical potential is given
by \be V_D^{phys}(R) = V_D(\mu, R) - V_D^{self}(\mu) \label{yh8}
\ee where both terms in the r.h.s. diverge if $\mu \to \infty$
while $V_D^{phys}(R)$ is well defined in this limit (see, e.g.
\cite{bp}). It is known that perturbative series for the static
potential in Yang-Mills theory does not exponentiate, contrary to
abelian case \cite{sf}. Nevertheless it is convenient to
exponentiate Wilson loop (\ref{wloopap}) as (see review
\cite{revour} and references therein) \be \lll W_D(C)\rrr = \exp
\>\left(\sum\limits_{m=1}^{\infty} i^m \>{\Delta}_D^{(m)}(R,T)
\right) \label{clusteree} \ee where each ${\Delta}_D^{(m)}(R,T)$
is equal to well defined linear combinations of gauge-invariant
field correlators (see (\ref{oioi}), (\ref{yh7}) below). We take
(\ref{potent1}) as a nonperturbative definition of the potential
in the present paper and problems of actual renormalization will
not be considered, correspondingly the argument $\mu$ is omitted
below. It is worth remembering however that all terms in the
series defined by (\ref{clusteree}) contribute to the potential
and are to be renormalized accordingly.

In perturbation theory each ${\Delta}_D^{(m)}(R,T)$ carries at
least a factor $g^m$ which is small, the full series however is
divergent due to factorial growth of the coefficients. On the
other hand, nonperturbatively there are no general reasons to
conclude that there exists some hierarchy between correlators with
different $m$. The general terminology calls the ensemble of the
correlators {\it stochastic} if $|{\Delta}_D^{(m)}(R,T)| \ge
|{\Delta}_D^{(m+1)}(R,T)|$ and {\it coherent}, if
$|{\Delta}_D^{(m)}(R,T)| \simeq |{\Delta}_D^{(m+1)}(R,T)|$. It can
be argued that the vacuum of Yang-Mills theory is stochastic and
not coherent \cite{ds1,ds2,ds3} and Casimir scaling is one of the
strongest arguments in favor of that \cite{sim1,ss1,ss2}. The
ultimate form of stochasticity is given by {\it Gaussian
dominance} scenario when one keeps only the lowest nontrivial
irreducible correlator in (\ref{clusteree}) corresponding to
$m=2$.
  The area law for the Wilson loop in the confining regime is given in the
Gaussian vacuum in the following form: \be {\Delta}_D^{(2)}(R,T) =
d_{D/f}\cdot \sigma_f^{(2)} \cdot R \cdot T \label{al} \ee where
$\sigma_f^{(2)} \simeq \lll \T F^2(0)\rrr T_g^2$ and typical gluon
correlation length $T_g$ is of the order of $0.2$ Fm for pure
$N=3$ gluodynamics (see, e.g. \cite{dp8,lat}). As it is clear from
(\ref{al}), Gaussian string tension $\sigma_D^{(2)}$ is always
proportional to the eigenvalue of quadratic Casimir operator for
the representation $D$ and therefore Gaussian dominance implies CS
(not vice versa, however). Therefore to address the question of
possible deviations from CS law, one has to go beyond bilocal
approximation. This will be done in the next, somewhat more
technical section.

\section{Deviations from CS law}

We are to study  the general structure of CS violating terms in
the cluster expansion. All calculations are performed for the
simplest case of flat rectangular contour $C=R\times T$ in
Euclidean space and its minimal surface. It does not mean any loss
of generality since all expressions below can be trivially
generalized to the case of arbitrary contour, whose minimal
surface has the disk topology. In the case under study it is
natural to adopt the so called contour gauge condition, introduced
in \cite{cont,cont1} as generalization of Fock-Schwinger fixed
point gauge \cite{fsch}. It is defined by the condition that the
phase factor orthogonal to the temporal axis is always equal to
unity, i.e. \be \Phi(\vec x , \vec y ) = \mbox{P}\exp \left[ i
(\vec y - \vec x)\int\limits_{0}^{1} {\vec A} (\vec x(1-s) + \vec
y s , z_4) ds \right] \equiv 1 \label{gauge1} \ee where $z_4 = x_4
=y_4$ and $A(x) = A^a(x)T_D^a$. Without loss of generality one may
also choose \be \Phi(x_4 , y_4) = \mbox{P}\exp \left[ i
\int\limits_{x_4}^{y_4} A_{4}(\vec 0, z_4) dz_{4} \right] \equiv 1
\label{gauge2} \ee The Wilson loop (\ref{wloopap}) takes
especially simple form in this gauge \be \lll W_D(C)\rrr =
\left\lll {\T}_D \>{\mbox P}\exp\left(i\int\limits_0^T
A_{4}^a(\vec z, t) T_D^a dt\right)\;\right\rrr \label{wloop1} \ee
where $| \vec z | =R$ and the potential $A_{4}^a(\vec z, t)$ is
expressible in terms of the electric components of the field
strength \be A_{4}^a(\vec z, t) = \int\limits_{0}^{1} ds \>\vec z
{\vec E}^a ( s \vec z ; t) \;\;\; ; \;\;\; \vec E^a_i(x) =
F^a_{i4}(x)
 \label{af} \ee
The contour gauges are well suited for studies of general
properties of the gauge theories, but present considerable
difficulties in perturbative calculations. Expanding
(\ref{wloop1}) one gets, in the matrix form \be \lll W_D(C) \rrr =
{\T}_D \left( {\hat{\bf 1}}_D + \sum\limits_{n=1}^{\infty}\> i^n
\> \int\limits_0^T dt_1 \> ... \int\limits_0^{t_{n-1}} dt_n \>
\lll A_4({\vec z}, t_n) ... A_4({\vec z}, t_1) \rrr \right)
\label{clusto} \ee Each  average in the r.h.s. of (\ref{clusto})
is proportional to the unit matrix in color space due to
triviality of the vacuum quantum numbers, so the whole expression
is, namely \be \int\limits_0^T dt_1 \> ... \int\limits_0^{t_{n-1}}
dt_n \> \lll A_4({\vec z}, t_n) ... A_4({\vec z}, t_1) \rrr =
{\hat{\bf 1}}_D \cdot \lll \T_D \> {\cal K}_D^{(n)}(R,T) \rrr
\label{kk} \ee
 With (\ref{gauge1}), (\ref{gauge2}) and (\ref{af}) at
hands one can easily rewrite every $\lll \T_D \> {\cal
K}_D^{(n)}(R,T) \rrr$ in manifestly gauge-invariant form (see
\cite{revour} and references therein).

We will omit arguments $R$ and $T$ in the correlators $\lll \T_D
\> {\cal K}_D^{(n)}(R,T) \rrr$ below for simplicity of notation.
We also do not discuss the actual dependence of each correlator on
its arguments in the present paper. Lattice data suggest to fit
the correlator as a sum of hard perturbative part and soft
nonperturbative one, the latter being responsible for confinement.
Notice that the series (\ref{clusto}) is not strictly speaking
perturbative series in the sense that each term in (\ref{clusto})
is given in perturbation theory by its own perturbative subseries.
It is of no importance for our analysis which will proceed in
terms of general expression and can be applied to whatever
dynamics if it provides the set of correlators which obey Eq.
(\ref{kk}). The perturbation theory itself can be regarded as a
good example.

It is now crucial to determine the relationship between $\lll \T_D
\> {\cal K}_D^{(n)} \rrr$ and ${\Delta}_D^{(m)}$. In the case of
interest $\lll \T_D {\cal K}_D^{(1)} \rrr \equiv 0$, so the first
term ${\Delta}_D^{(1)}$ also vanishes while the relation between
higher correlators is to be read off from the Taylor expansion of
(\ref{clusteree}): \be {\Delta}_D^{(2,3)} = \lll \T_D {\cal
K}_D^{(2,3)} \rrr \;\; ;
 \;\; {\Delta}_D^{(4)} = \lll \T_D {\cal K}_D^{(4)} \rrr - \frac12 \>
\left[ \lll \T_D {\cal K}_D^{(2)} \rrr \right]^2  \label{oioi} \ee
and the terms of higher orders are built in full analogy with
(\ref{oioi}): \be {\Delta}_D^{(m)} = \lll \T_D {\cal K}_D^{(m)}
\rrr - (...) \label{yh7} \ee where $(...)$ stays for the sum of
products of the correlators $\lll \T_D {\cal K}_D^{(p)} \rrr$ of
lower orders, $p<m$. The expression (\ref{yh7}) does not represent
the Green's function of any physical colorless state, instead, it
describes the propagation of some gluelump state, which is not a
physical excitation in gluodynamics. It is also worth saying that
(\ref{oioi}), (\ref{yh7}) should be considered as a {\it
definition}, valid beyond perturbation theory (each correlator is
given by the corresponding perturbative series in the latter).

The potential $V_D(R)$ is given up to an additive constant by \be
V_D(R) = - \lim_{T\to\infty} \> \frac{1}{T} \>
\sum\limits_{m=2}^{\infty} i^m \>{\Delta}_D^{(m)} \label{potent}
\ee   We are going to study now the structure of special linear
combination of potentials, measuring the deviations from Casimir
scaling  \be \delta V_{D_2 / D_1} = V_{D_2}(R) - d_{D_2 / D_1}
\cdot V_{D_1}(R) \label{deltav} \ee We confine our attention to
the case of fundamental and adjoint representations in this paper,
in this particular case (\ref{deltav}) takes the form \be \delta
V_{a/f} = - \lim_{T\to\infty} \> \frac{1}{T} \>
\sum\limits_{m=2}^{\infty} i^m \>\left( {\Delta}_{adj}^{(m)} -
\frac{2N^2}{N^2 -1} \cdot {\Delta}_{fund}^{(m)} \right) = -
\lim_{T\to\infty} \> \frac{1}{T} \> \sum\limits_{m=2}^{\infty} i^m
\delta \Delta_{a/f}^{(m)}
 \label{deltav1}
\ee As a next step, we are exploiting the identity \be \left\lll
\left| W_{f}(C)\right|^2 \right\rrr = \frac{1}{N^2} +
\left(1-\frac{1}{N^2} \right)\lll W_{a}(C)\rrr \label{wfa} \ee We
need to compute the quantity $\left\lll \left| W_{f}(C)\right|^2
\right\rrr$ in terms of correlators (\ref{kk}). It can be done as
follows. The average of the product is given by
$$ \lel |W_{f}(C)|^2 \rir = {\T}_{12} \lel
\>{\mbox P}\exp\left(i\int\limits_0^T A_{4}^a(\vec z, s) t^a
ds\right) \>{\mbox P}\exp\left(-i\int\limits_0^T A_{4}^a (\vec z,
\tau ) t^a d\tau \right) \rir  =
$$
\be = {\T}_{12} \lel \left( {\bf{\hat 1}}_{f_1} +
\sum\limits_{n=1}^{\infty} i^n {\cal K}_{f_1}^{(n)} \right)\cdot
\left( {\bf{\hat 1}}_{f_2} + \sum\limits_{p=1}^{\infty} (-i)^p
{\cal K}_{f_2}^{(p)} \right) \rir \label{wet1} \ee where the trace
$\T_{12} = \T_{f_1} \T_{f_2}$ acts on the color indices
corresponding to the first and the second loop independently.
Global gauge invariance dictates the following form of the
corresponding averages (see also \cite{wephrev}): $$ \lll {\cal
K}_{f}^{(n)} \rrr = {\bf{\hat 1}}_{f} \cdot \lll \T_f {\cal
K}_{f}^{(n)} \rrr \;\;\; ; \;\;\;
 \lll {\cal K}_{f_1}^{(n)} \> {\cal K}_{f_2}^{(p)} \rrr =
$$
$$
= {\bf{\hat 1}} \cdot \frac{1}{N^2 -1}\> \left( N^2 \> \lll \T_{f}
{\cal K}_{f_1}^{(n)} \;\T_{f} {\cal K}_{f_2}^{(p)} \rrr - \lll
\T_{f} \> {\cal K}_{f_1}^{(n)} \> {\cal K}_{f_2}^{(p)} \rrr
\right) +
$$
\be + \> {\bf{\hat e}} \cdot \frac{N}{N^2 -1} \> \left( \lll
\T_{f} \> {\cal K}_{f_1}^{(n)} \> {\cal K}_{f_2}^{(p)} \rrr - \lll
\T_{f} {\cal K}_{f_1}^{(n)} \; \T_{f} {\cal K}_{f_2}^{(p)} \rrr
\right) \label{srd} \ee The matrices ${\bf{\hat 1}}_f$, $\bf{\hat
1}$ and $\bf{\hat e}$ can be written in index notation as \be
[{\bf{\hat 1}}_f]_{\alpha_1\beta_1} = \delta_{\alpha_1 \beta_1}
\;\; ; \;\; [{\bf{\hat 1}}]_{\alpha_1\beta_1; \alpha_2\beta_2} =
\delta_{\alpha_1 \beta_1} \delta_{\alpha_2 \beta_2} \;\; ; \;\;
[{\bf{\hat e}}]_{\alpha_1\beta_1; \alpha_2\beta_2} =
\delta_{\alpha_1 \beta_2} \delta_{\alpha_2 \beta_1} \label{e21}
\ee where $\alpha_{1,2} , \beta_{1,2} $ are fundamental indices
running from $1$ to $N$. The algebra they obey encodes the effects
of path ordering. In our case since the orientations of minimal
surfaces corresponding to the first and the second multiplier in
(\ref{wet1}) are antiparallel, the matrix $\bf{\hat e}$ should be
multiplied from the right with respect to indices carrying
subscript "2" but from the left with respect to indices carrying
subscript "1". It results in the following algebra: \be {\bf{\hat
e}} \cdot {\bf{\hat 1}} = {\bf{\hat 1}} \cdot {\bf{\hat e}} =
{\bf{\hat e}} \; ; \; {\bf{\hat 1}}^2 = {\bf{\hat 1}} \; ; \;
{{\bf{\hat e}}}^2 = N {\bf{\hat e}} \label{e26} \ee and the traces
are given by \be \T_{12} \> {\bf{\hat 1}} = 1 \;\;\; ; \;\;\;
\T_{12} \> {\bf\hat e} = \frac{1}{N} \ee It is convenient to
represent (\ref{wet1}) in the following form:
$$ \lel W_{f}(C) W_{f}^{\dagger}(C) \rir = {\T}_{12} \exp\left(
{\bf{\hat 1}} \cdot A + {\bf{\hat e}} \cdot B \right) =  {\T}_{12}
\left( {\bf{\hat 1}}\cdot \exp A + \right.
$$
\be + \left.
 \frac{{\bf{\hat e}}}{N} \cdot \left(
\exp(A+BN) - \exp A \right)\right) = \frac{1}{N^2}\exp(A+BN) +
\left(1-\frac{1}{N^2}\right) \exp A \label{f87}
 \ee
Comparing (\ref{f87}) with (\ref{wfa}), (\ref{wet1}) and
(\ref{srd}) one concludes that
$$
\exp(A+BN) \equiv 1
$$
$$ \exp A = \lll W_{a}(C)\rrr = 1 + \sum\limits_{n=1}^{\infty} i^n \lll \T_f {\cal
K}_{f}^{(n)}\rrr +  \sum\limits_{p=1}^{\infty} (-i)^p \lll \T_f
{\cal K}_{f}^{(p)} \rrr
 +
 $$
 \be
 +
\sum\limits_{n=1}^{\infty} \sum\limits_{p=1}^{\infty} \frac{i^n
(-i)^p}{N^2 -1} \left( N^2 \> \lll \T_{f} {\cal K}_{f_1}^{(n)}
\;\T_{f} {\cal K}_{f_2}^{(p)} \rrr - \lll \T_{f} \> {\cal
K}_{f_1}^{(n)} \> {\cal K}_{f_2}^{(p)} \rrr \right) \label{erk}
\ee This is exact expression for adjoint Wilson loop average,
written in terms of fundamental correlators and traces and as such
it is suitable for our analysis. We are to study the difference
\be \delta V_{a/f} = - \lim_{T\to\infty} \frac{1}{T} \left( \log
\lll W_a(C) \rrr - \frac{2N^2}{N^2 -1} \log \lll W_f(C) \rrr
\right) \label{op4} \ee where the first term in the curly brackets
in (\ref{op4}) is equal to $\log \lll W_a(C) \rrr = A$ where $A$
is defined by (\ref{erk}), while the second term is given by
formal series \be\log \lll W_f(C) \rrr = \log \left(1 +
\sum\limits_{n=1}^{\infty} i^n \lll \T_f {\cal K}_{f}^{(n)} \rrr
\right) =  - \sum\limits_{k=1}^{\infty} \frac{(-1)^{k}}{k} \>
\left[ \sum\limits_{n=1}^{\infty} i^n \lll \T_f {\cal K}_{f}^{(n)}
\rrr \right]^k \label{fun3} \ee To proceed further one can exploit
the identity $\left\lll [W_f(C)]_{\alpha\beta}
[W_f^{\dagger}(C)]_{\beta\alpha} \right\rrr \equiv 1$ or, in terms
of correlators \be \sum\limits_{n=1}^{\infty} i^n \lll \T_f {\cal
K}_{f}^{(n)}\rrr + \sum\limits_{p=1}^{\infty} (-i)^p \lll \T_f
{\cal K}_{f}^{(p)} \rrr \equiv
 - \sum\limits_{n=1}^{\infty} \sum\limits_{p=1}^{\infty} {i^n
(-i)^p} \lll \T_{f} \> {\cal K}_{f_1}^{(n)} \> {\cal
K}_{f_2}^{(p)} \rrr \label{erk2} \ee It is now straightforward to
rewrite (\ref{erk}) as:
$$ \log \lll W_a(C) \rrr = A = \log \left( 1 + \frac{N^2}{N^2 -1}\left(
\sum\limits_{n=1}^{\infty} i^n \lll \T_f {\cal K}_{f}^{(n)}\rrr +
\sum\limits_{p=1}^{\infty} (-i)^p \lll \T_f {\cal K}_{f}^{(p)}
\rrr \right.\right. +
$$
$$ + \left.\left.\vphantom{\frac{N^2}{N^2 -1}
\sum\limits_{n=1}^{\infty}} \sum\limits_{m=2}^{\infty}
\sum\limits_{p=1}^{m-1} i^m (-1)^p \lll \T_{f} {\cal
K}_{f_1}^{(m-p)} \;\T_{f} {\cal K}_{f_2}^{(p)} \rrr \right)\right)
=
$$
$$ = - \sum\limits_{k=1}^{\infty} \frac{(-1)^{k}}{k} \> \left(
\frac{N^2}{N^2 -1} \right)^k \sum_{l=0}^k C_k^l \left[
\sum\limits_{n=1}^{\infty} i^n \lll \T_f {\cal K}_{f}^{(n)}\rrr +
\sum\limits_{p=1}^{\infty} (-i)^p \lll \T_f {\cal K}_{f}^{(p)}
\rrr\right]^l \cdot $$ \be \cdot \left[ \sum\limits_{m=2}^{\infty}
i^m \sum\limits_{p=1}^{m-1} (-1)^p \lll \T_{f} {\cal
K}_{f_1}^{(m-p)} \;\T_{f} {\cal K}_{f_2}^{(p)} \rrr \right]^{k-l}
\label{uuok} \ee

Till now we have not made any approximations yet and the
expressions (\ref{fun3}), (\ref{uuok}) are valid for arbitrary
contour $C$ and any contour gauge. It is worth reminding that
despite each term in (\ref{fun3}), (\ref{uuok}) is
gauge-invariant, its actual value depends on the profile of the
contours used in (\ref{gauge1}), (\ref{gauge2}) but this
dependence is cancelled after summation of all terms. Likewise the
property of  path-dependence of Gaussian correlator (found in
\cite{digmeg} to be rather strong) shed no light on the problem of
CS since gauge-invariant path-dependent bilocal correlator $ \lll
\T_D F_{\mu\nu}(x,x_0)F_{\rho\sigma}(y,x_0)\rrr $ where $
F_{\mu\nu}(x,x_0) = \Phi(x_0,x) F_{\mu\nu}(x)
[\Phi(x_0,x)]^{\dagger}$ and $\Phi(x_0,x)$ are the phase factors
(chosen to be unit matrices in our analysis) is always
proportional to $C_D$ for any path, taken in $\Phi(x_0,x)$.

We can now use the Euclidean rotational 4-invariance of our
problem which dictates $ \lll W_f(C) \rrr = \lll W_f^{\dagger}(C)
\rrr $ or, at the level of correlators, \be
\sum\limits_{n=1}^{\infty} i^n \lll \T_f {\cal K}_{f}^{(n)}\rrr =
\sum\limits_{p=1}^{\infty} (-i)^p \lll \T_f {\cal K}_{f}^{(p)}
\rrr\label{o8} \ee Let us now look at the general structure of
terms in $\delta V_{a/f}$. It is straightforward to notice that
each $\delta\Delta_{a/f}^{(m)}$ gets contributions from
irreducible correlators of gauge-invariant operators of the type $
\lll \T_{f} {\cal K}_{f_1}^{(m-p)} \;\T_{f} {\cal K}_{f_2}^{(p)}
\rrr $. The contribution from the single correlator of the same
order $ \lll \T_{f} {\cal K}_{f}^{(m)} \rrr $ can be shown to be
exactly cancelled in (\ref{op4}) (this cancellation happens in the
terms corresponding to $k=1$ in (\ref{fun3}) and (\ref{uuok})). To
illustrate the point, consider the lowest nontrivial order of
cluster expansion violating CS, which corresponds to $m=4$. One
gets \be \delta\Delta_{a/f}^{(4)} = \left(\frac{N^2}{N^2 -1}
\right)\cdot\left( \lll \T_{f_1} {\cal K}_{f_1}^{(2)} \T_{f_2}
{\cal K}_{f_2}^{(2)} \rrr - \left( 1 + \frac{2}{N^2 -1} \right)
\left[ \lll \T_f {\cal K}_f^{(2)} \rrr \right]^2 \right)
\label{f5} \ee  This expression is gauge-invariant and the fields
$A_4(x)$ in the correlators $\lll \T {\cal K}^{(n)} \rrr $ are
functionals of the field strengths $F_{i4}(x)$ as given by
(\ref{af}).

The l.h.s. of expression (\ref{f5}) vanishes identically if CS law
is exact. Posed differently, it measures the deviation from CS in
terms of integral moments of particular correlators, entering the
r.h.s. of (\ref{f5}). The coefficient in front of the second term
in the r.h.s. of (\ref{f5}) (equal to $5/4$ in the case of
$SU(3)$) provides exact cancellation of reducible contributions
from the first term, so both sides of equality (\ref{f5}) have
area law asymptotics in confinement regime. In large $N$ limit
they both vanish, as it should be. The same pattern holds in
higher orders of cluster expansion. Casimir scaling violating
contributions are given at fixed $m$ by the irreducible
correlators of colorless operators $\lll \T_{f} {\cal
K}_{f_1}^{(m-p)} \;\T_{f} {\cal K}_{f_2}^{(p)} \rrr $.  It is to
be compared with the expressions (\ref{oioi}), (\ref{yh7}) where
the potential itself is written in terms of correlators $\lll
\T_{f} {\cal K}_{f}^{(m)} \rrr $.

\section{Conclusion}

Expression (\ref{f5}) and analogous formulas for higher orders of
cluster expansion formalize the statement made in \cite{ss2} that
violations of CS could be a result of exchange by colorless
states. We have considered only the case of adjoint and
fundamental potentials in the present paper but no reasons for
breaking of this mechanism in other representations are seen. The
qualitative picture corresponds to the confining string worldsheet
populated by irreducible correlators like $\lll \T_D \> {\cal
K}_D^{(2)} \rrr$ and correlators of higher orders made of
gauge-invariant expectation values of path-dependent field
strength operators. On the minimal surface the whole ensemble
contributes to the total string tension in such a way that
Gaussian contribution is numerically dominant, as CS suggests. In
other words, CS indicates that this ensemble is quasi-free in the
sense that interactions between such gauge-invariant objects as
$\T F F$ mediated by glueball states and destroying CS are rather
weak. Phenomenologically, we expect the r.h.s. of expressions like
(\ref{f5}) to be suppressed by some power of inverse glueball
mass. The actual smallness of deviations from CS as observed on
the lattice simply follows in this case from the fact that even
the lightest glueball is rather heavy, of the order of 1.5 GeV.
Moreover, one can have a glimpse of general reason why many
microscopic confinement models fail to reproduce CS. To this end,
the model, apart from explaining string tension formation and area
law of the Wilson loop at intermediate distances for all
representations ("CS plateau") must also describe the phenomenon
of mass gap itself and come out with the correct (rather large)
masses of colorless states. One can mention, on the other hand,
that reasonable prediction of the instanton liquid model for the
mass of the lowest glueball $(1.4 \pm 0.2 )$ GeV \cite{shuryak} is
in some contradiction with the strong violations of CS for the
static potential in this model \cite{ss2}. This paradox is
possibly resolved by the fact that the instanton-antiinstanton
ensemble is coherent and the cluster expansion is not well suited
for analysis in this case.

\newpage

{\bf \large Acknowledgments }

\bigskip

The discussions with Yu.A.Simonov are gratefully acknowledged. The
author is thankful to the foundation "Fundamenteel Onderzoek der
Materie" (FOM), which is financially
 supported by the Dutch National Science Foundation (NWO).
 He also acknowledges the support from the grants
RFFI-00-02-17836, RFFI-01-02-06284 and from the grant INTAS
00-110.

\bigskip

%%%%%%%%%%%%%%%%%%%%%%%%%%%%%%
%  BIBLIOGRAPHY
%%%%%%%%%%%%%%%%%%%%%%%%%%%%%%

\end{document}